\documentclass[a4paper]{jpconf}
\usepackage{graphicx}
\begin{document}
\title{Wide-field solar adaptive optics in a layer-oriented approach}

\author{Agla\'e Kellerer}

\address{University of Cambridge, 
Cavendish Laboratory, 
JJ Thomson Av., 
Cambridge, CB30HE, UK}

\ead{ak935@cam.ac.uk}

\begin{abstract}

We discuss a layer-oriented approach to multi-conjugate adaptive optics (MCAO) in solar imaging. 
The technique is a complement to the current star-oriented MCAO and appears as a necessary alternative when large field sizes are desired in solar observations. The basic procedure of the layer oriented method is indicated, and its characteristics are then illustrated in terms of numerical simulations. 

\end{abstract}

\section{Introduction}

To obtain wide-field images of the solar surface one corrects the atmospheric turbulence through {\it multi-conjugate adaptive optics\/}. MCAO uses several deformable mirrors to correct separately the phase distortions caused by turbulence in atmospheric layers at different altitudes. To this purpose one needs to estimate the phase distortions produced at these different altitudes. This information is obtained by using not just one wavefront sensor -- as in conventional AO -- but by tomographic evaluation of the phase profiles measured by several wavefront sensors.

The established method in solar MCAO uses the {\it star-oriented\/} approach (SO) \cite{Berkefeld, Montilla} where the wavefront distortions are sensed across the telescope pupil for, say, 19 directions distributed over the field of view. The 19 phase profiles permit then a tomographic estimation of the phase fluctuations caused in the different layers conjugated to the deformable mirrors. Solar AO uses Shack-Hartmann (SH) wavefront sensors, where each SH lenslet images an element of the solar surface, rather than an unresolved star in nighttime AO. The field width must, therefore, be somewhat larger for each lenslet, which means that the effective area sensed by it increases sufficiently at high layers for the signal to be slightly washed out\,\cite{Bechet}. While the weakened signal from higher altitudes is an undesirable feature in the star-oriented method, the tomographic method does work adequately, as long as the field width remains reasonably narrow. The correlation of solar SH images requires typically $20\times 20$ pixels of angular resolution $0.4''$. With this $8''$ viewing angle, a 0.10\,m diameter sub-aperture covers then a 0.41\,m diameter surface at 8\,km.

Yet one may consider the opposite case where -- on purpose -- the viewing angle of each lenslet is made as wide as the total viewing angle. If conjugated to the telescope pupil, the sensor then registers predominantly the phase distortions caused nearby, while the signal from higher altitudes is largely washed out. This suggests a fairly straight-forward alternative to the {\it star-oriented\/} approach. Using a few sensors conjugated to the most relevant altitudes one would have a method that conveniently provides the information for different layers directly. More importantly the method works best for the largest fields of view, because such fields permit the most effective zeroing-in on the specified layer. This {\it layer-oriented\/} approach (LO) has, in fact, been introduced successfully for night-time MCAO\,\cite{Ragazzoni, Diolaiti, Verinaud}. For solar observations it has not been utilized, and it will require different procedures there, but we propose that the technique be introduced as useful complement to the current SO-MCAO in solar imaging and, in fact, as a necessary alternative when large field sizes are desired in solar observations\,\cite{LO, LO2}. First the basic procedure will be indicated, then the characteristics of this layer oriented method for solar observations will be illustrated in terms of a few simulations.

\section{Principle of solar layer-oriented MCAO}

In the star-oriented approach each sensor assesses, across the telescope pupil, the phase shift of the photons incident in the specified direction. In the layer-oriented approach each sensor assesses, across the conjugate altitude, the phase shift of the photons incident from the entire field of view. Since the fluctuations from other altitudes tend to average out, the sensor registers predominantly the fluctuations due to the conjugated altitude. Night-time layer-oriented systems employ for this purpose pyramid sensors\,\cite{Ragazzoni, Diolaiti, Verinaud}. Each pyramid collects the light from one guide star, and the light from the different stars is then combined on a detector optically conjugated to the specified altitude. This is not a suitable solution in solar AO, where one deals with an extended source rather than multiple point sources. We have therefore suggested to implement the layer-oriented approach with SH sensors\,\cite{LO, LO2}. The method is now briefly described.

\begin{figure}[h]
\includegraphics[width=.3\textwidth]{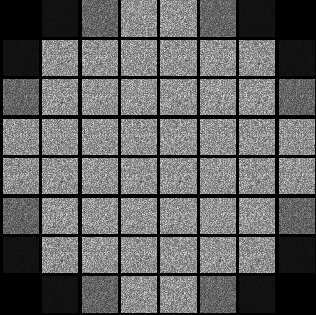}\hspace{2pc}
\includegraphics[width=.3\textwidth]{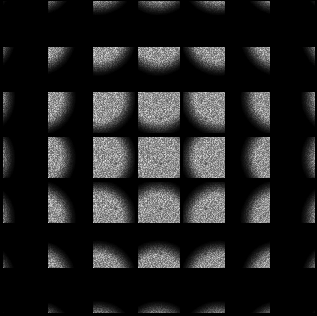}\hspace{2pc}
\includegraphics[width=.3\textwidth]{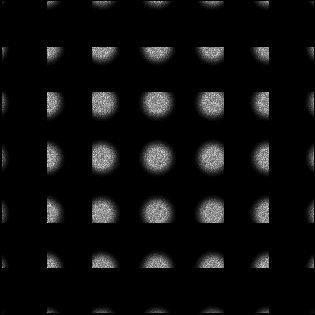}
\caption{\label{fig:setup} In a layer-oriented MCAO system the lenslet array of a SH sensor is conjugated to a turbulent layer (left, middle, right panels: 0\,km, 4\,km and 10\,km) and the sensor registers predominantly the phase fluctuations produced at that altitude. }
\end{figure}

In conventional AO the lenslet array of a SH sensor is conjugated to the telescope pupil, and the detector in the focal plane of the array registers for each lenslet an image of the observed solar area. In the layer-oriented approach, additional  lenslet arrays are conjugated to dominant turbulent layers.  Each lenslet images then only those points of the solar area for which some of the emitted photons traverse on their way to the telescope this sub-pupil. Fig.\,\ref{fig:setup}  depicts for three different altitudes the images registered behind each lenslet. 

The vignetting of the lenslet images is no artefact of the layer oriented method, it merely reflects the fact that -- at the higher altitudes -- photons from only part of the observed solar area traverse the particular sub-pupil on their way to the telescope pupil; merely these photons experience the atmospheric turbulence at this location and require the corresponding phase correction.

Consider a simplified scenario where the turbulence is entirely located at the conjugate height of a SH sensor: wavefronts from all angular directions are affected by the turbulence similarly, i.e. the phase fluctuations are independent of the field direction and the image behind a lenslet is therefore globally shifted. If, on the contrary, the turbulence is located far from the conjugate height, then the fluctuations vary with field direction and the SH images are distorted rather than shifted. In the layer-oriented approach, the image shift forms the sensor-signal; it is an approximation of close-by turbulence and is used to shape the deformable mirror that is conjugated to the same height as the sensor. 

As an aside we note that the wavefront distortions are sensed continuously over the entire field of view: the correction loop therefore automatically converges to an optimal solution for the entire field of view. This contrasts with the star-oriented approach where the correction for the entire field is extrapolated from measurements along multiple discrete directions.

\section{Reduction of the field-of-view to the relevant photon fluence}\label{sec:sign}

\begin{figure}[h]
\includegraphics[width=0.5\textwidth]{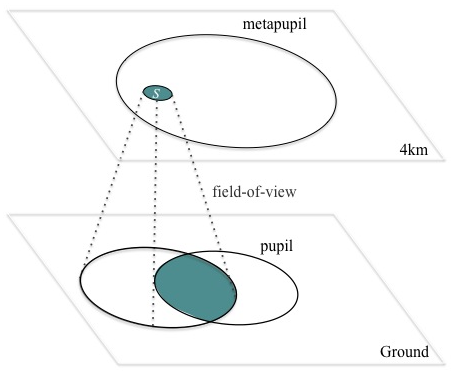}\hspace{2pc}%
\begin{minipage}[b]{14pc}\caption{\label{fig:vign1} Solar layer-oriented MCAO: Atmospheric fluctuations at $S$ affect only part of the field-of-view. A SH lenslet conjugated to surface $S$ characterizes, at the ground, the phase fluctuations in the intersection between the pupil and the disc traced-out by the field of view.}
\end{minipage}
\end{figure}

As shown on Fig.\,\ref{fig:setup}, SH lenslets that are conjugated above the pupil form vignetted images. Assume that the turbulence is entirely located in two layers: at the ground and 4\,km. One aims at determining the phase fluctuations introduced over a small surface $S$ located at the border of the 4\,km metapupil:  see Fig.\,\ref{fig:vign1}. 

In the solar layer-oriented approach, a SH lenslet is optically conjugated to surface $S$. The signal measured behind this lenslet corresponds to the phase gradient introduced over surface $S$ plus the gradient over a larger surface at the ground. In Fig.\,\ref{fig:vign1} this larger surface corresponds to the intersection between the pupil and the disc traced out by the field-of view: this is because only the photons that end up in the intersection with the pupil (i.e. are accepted by the telescope in normal operation) traverse  $S$. 
It has been suggested that the layer-oriented method is inoperable because the field-reduction restricts the effective surfaces at mis-conjugated altitudes, and thereby weakens the attenuation of mis-conjugated layers\,\cite{MW}.  
In fact field reduction affects any MCAO system, i.e. the star- and layer-oriented approaches in their night- and daytime implementation. It actually affects existing MCAO approaches more strongly than the proposed solar layer-oriented method. Field-reduction is therefore not a limitation for solar layer-oriented MCAO.

\begin{figure}[h]
\includegraphics[width=0.5\textwidth]{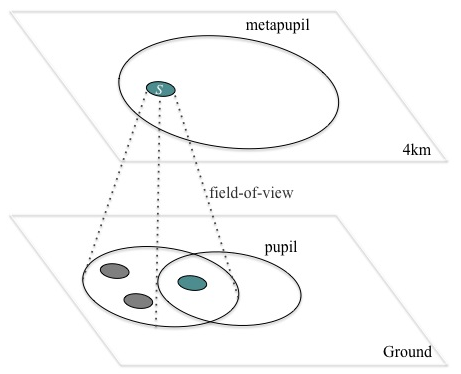}\hspace{2pc}%
\begin{minipage}[b]{14pc}\caption{\label{fig:vign3}Nighttime layer-oriented MCAO: For the sensor elements conjugated to surface $S$ the signal from the ground layer is not attenuated because only one star lies within the unvignetted field-of-view. }
\end{minipage}
\end{figure}

In a nighttime layer-oriented system, a detector -- rather than a lenslet array -- is optically conjugated to a turbulent layer. The fundamental principle of both layer-oriented methods are, however, identical: signals from mis-conjugated layers are attenuated because the corresponding phase fluctuations are averaged over larger effective surfaces. 
In the nighttime implementation, the averaging is done over the footprints of multiple guide stars. 
Fig.\,\ref{fig:vign3} assumes three guide stars. For the detector pixels conjugated to surface $S$ the signal from the ground layer is not attenuated because only one star lies within the unvignetted field-of-view. 
This example illustrates that field-reduction has more drastic consequences in nighttime than in solar layer-oriented MCAO: indeed, because of the continuous solar fields, mis-conjugated layers are always somewhat attenuated in the solar application.
Since the layer-oriented approach has been successfully implemented in nighttime MCAO, field-reduction can also not be a limitation for the solar application.

\section{Simulations}

Some numerical simulations in terms of {\it Yao\/}, an adaptive optical simulation code written by F. Rigaut under {\it Yorick\/} can illustrate the proposed method. We have written additional functions under {\it Yao\/} in order to simulate solar observations and wide-field Shack-Hartmann sensors conjugated above the telescope pupil.

Assume three sensor-mirror pairs conjugated to 0\,km, 4\,km and 10\,km. Each sensor propagates a $100''$ diameter field. The telescope diameter equals 1.6\,m. 
The signal measured by a sensor is exclusively used to activate its associated mirror: in other words, once the sensor signals are determined, the tomographic reconstruction is done.

\begin{figure}[h]
\includegraphics[width=\textwidth]{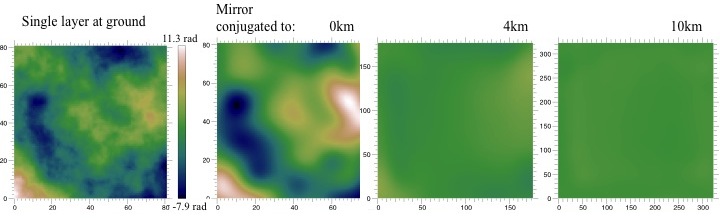}
\caption{\label{fig:sim1}Effect of a single layer at ground level: After one iteration, the mirror at ground level reproduces the shape of the phase screen while the two other mirrors remain almost flat. The system has assimilated that the turbulence is located at ground level.}
\end{figure}

Fig.\,\ref{fig:sim1} shows the effect of a single layer at ground level. The Fried parameter is set to $r_0=0.15$\,m. As expected, the turbulence is mainly detected by the sensor conjugated to the ground, while the two other sensors measure a negligible signal: After one iteration, the mirror conjugated to the ground reproduces the shape of the turbulent screen while the two other mirrors remain nearly flat. The system has assimilated that the turbulence is introduced at the ground -- the tomographic reconstruction is successful.

\begin{figure}[h]
\includegraphics[width=\textwidth]{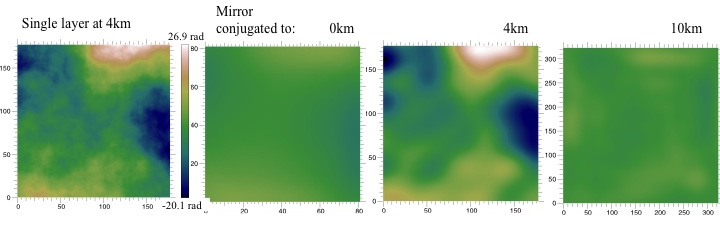}
\caption{\label{fig:sim2}Effect of a single layer at 4\,km: After one iteration, the mirror conjugated to 4\,km replicates the shape of the phase screen while the two other mirrors remain almost flat. The tomographic reconstruction is correct. }
\end{figure}

Fig.\,\ref{fig:sim2} represents the effect of a single layer at 4\,km with Fried parameter $r_0=0.15$\,m. The system again responds correctly: the mirror conjugated to 4\,km takes up the shape of the phase screen, while the two other mirrors remain approximately flat.

\begin{figure}[h]
\begin{center}
\includegraphics[width=\textwidth]{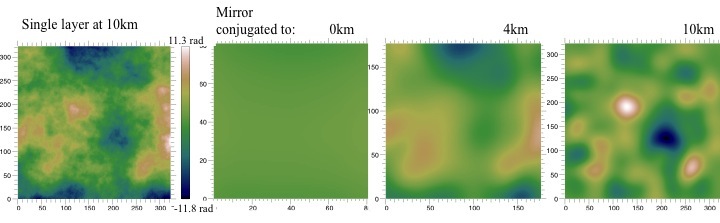}
\caption{Effect of a single layer at 10\,km. After one iteration, the 10\,km mirror reproduces the shape of the phase screen, while the 4\,km mirror replicates its lowest spatial frequencies: field-reduction has made the attenuation of mis-conjugated layers less efficient. This effect may be minimized by filtering out low spatial modes in the command matrix.  }
\label{fig:sim3}
\end{center}
\end{figure}

The effect of a single turbulent layer at 10\,km with Fried parameter $r_0=0.50$\,m is illustrated on Fig.\,\ref{fig:sim3}. After one iteration, the mirror conjugated to 10\,km replicates the phase screen, but the 4\,km mirror likewise reproduces the lowest spatial frequencies of the screen: The sensor conjugated to 4\,km has measured some signal -- field-reduction has made the attenuation of distant layers less efficient. After a few iterations however, the mirror conjugated to 4\,km flattens out because the sensor-mirror pair at 10\,km is more efficient at correcting the phase fluctuations.

\section{Conclusion}

Table \ref{tab:sum} summarizes differences between star- and layer-oriented solar MCAO correction.
A complication of the layer-oriented approach is that it requires rapid cross-correlations of large images. Further, it necessitates $\sim2000\times 2000$ pixel detectors that can be read-out above 1000\,Hz. The availability of such detectors is currently the main challenge for the solar layer-oriented approach.

\begin{table*}[!h]
\begin{center}
\begin{tabular}{| p{.2\textwidth} p{.35\textwidth} p{.35\textwidth}|} \hline
& Star-oriented & Layer-oriented \\ \hline
High-turbulence & 
..is attenuated & 
{\bf Each sensor is most sensitive to its associated turbulent layer\/} \\\hline
Spatial and temporal sampling & 
..tuned to integral turbulence & 
{\bf ..tuned to associated layer\/} \\\hline
Tomographic reconstruction & 
..is done computationally & 
{\bf ..is done optically\/} \\
&
..becomes more complex with increasing field size & 
{\bf ..benefits from large field sizes\/} \\\hline
Cross-correlation & 
{\bf ..over $\sim$ 20$\times$20 pixels\/} & 
..over $\sim 200\times200$ pixels:
computationally intensive \\\hline
Detectors & 
{\bf Small detectors / Large detectors with ROI read-out\/} &
Large detectors with high read-out frequencies \\\hline
\end{tabular} 
{\caption{Comparison between star- and layer-oriented solar MCAO. Advantages in bold.}}
\label{tab:sum}
\end{center}
\end{table*} 

\section*{Bibliography}

\bibliographystyle{unsrt}
\bibliography{Ref}

\end{document}